\documentclass[pra,twocolumn,showpacs,superscriptaddress]{revtex4}
\usepackage{amsmath,amsfonts,algorithm,algorithmic}

\usepackage{bm}
\usepackage{ifpdf}
  \usepackage[pdftex]{graphicx}
  \graphicspath{{figures/pdf/}}
  \DeclareGraphicsExtensions{.pdf}
\def\vec#1{{\bf#1}}
\def\uvec#1{\hat{\bf#1}}

\def\op#1{#1}
\def\ket#1{| #1 \rangle}
\def\bra#1{\langle #1 |}
\def\ip#1#2{\langle #1 \mid #2 \rangle}

\def\ave#1{\langle #1 \rangle}
\def\norm#1{\| #1 \|}
\def\ceil#1{\lceil #1 \rceil}
\def\floor#1{\lfloor #1 \rfloor}

\def\aa{\alpha}

\def\Tr{\operatorname{Tr}}

\def\diag{\operatorname{diag}}

\def\su{\mathfrak{su}}
\def\SU{\mathbb{SU}}
\def\SO{\mathbb{SO}}

\def\RR{\mathbb{R}}

\def\E{\mathcal{E}}
\def\F{\mathcal{F}}
\def\eps{\epsilon}
\def\sx{\op{\sigma}_x}
\def\sy{\op{\sigma}_y}
\def\sz{\op{\sigma}_z}

\def\Id{\mathbb{I}}
\def\Rh{R_{\uvec{h}}}
\def\Rg{R_{\uvec{g}}}
\def\Rx{R_{\uvec{x}}}

\def\Rz{R_{\uvec{z}}}

\def\QUAD{\hspace{2.5em}}
\begin{document}
\title{Improving quantum gate fidelities using optimized Euler angles}
\author{K.~Ch.~Chatzisavvas}\email{kchatz@auth.gr}
\affiliation{Physics Department, Aristotle University of Thessaloniki,
             54124 Thessloniki, Greece}
\author{G.~Chadzitaskos}\email{goce.chadzitaskos@fjfi.cvut.cz}
\affiliation{Department of Physics, FNSPE, Czech Technical University,
             Brehova 7, CZ-115 19 Praha 1, Czech Republic}
\author{C.~Daskaloyannis}\email{daskalo@math.auth.gr}
\affiliation{Mathematics Department, Aristotle University of
             Thessaloniki, 54124 Thessloniki, Greece}
\author{S.~G.~Schirmer}\email{sgs29@cam.ac.uk}
\affiliation{Department of Applied Maths and Theoretical Physics,
             University of Cambridge, Wilberforce Road, Cambridge, CB3
         0WA, United Kingdom}
\affiliation{Department of Maths and Statistics,
             University of Kuopio, PO Box 1627, 70211 Kuopio, Finland}
\date{\today}

\begin{abstract}
An explicit algorithm for calculating the optimized Euler angles for
both qubit state transfer and gate engineering given two \emph{arbitary}
fixed Hamiltonians is presented.  It is shown how the algorithm enables
us to efficiently implement single qubit gates even if the control is
severely restricted and the experimentally accessible Hamiltonians are
far from orthogonal.  It is further shown that using the optimized Euler
angles can significantly improve the fidelity of quantum operations even
for systems where the experimentally accessible Hamiltonians are nearly
orthogonal.  Unlike schemes such as composite pulses, the proposed
scheme does not significantly increase the number of local operations or
gate operation times.
\end{abstract}

 \maketitle

\section{Introduction}
\label{sec:introduction}

Quantum computing~\cite{00Nielsen} generally relies on the decomposition
of arbitrary multi-qubit operations into products of elementary single
and two-qubit gates, which must be implemented with very high
fidelity. Although the availability of an entangling two-qubit gate is
crucial for universal quantum computing~\cite{PRS449p669}, single qubit
operations dominate virtually any decomposition of a multi-qubit quantum
gate.  For example, if we decompose a two-qubit gate into two-qubit
gates that can be generated by a natural Ising interaction and local
operations using the Cartan decomposition~\cite{Knapp02}, at most three
two-qubit terms are required in addition to $12$ single qubit rotations.
Therefore the fidelity of single qubit gates is critical, as even small
single qubit gate errors quickly accumulate, resulting in poor
multi-qubit gate fidelities even if the entangling gate is perfect.

One approach to improving gate fidelities and gate operation times is
using optimal control.  In general, optimal control fields can be
derived by simultaneous optimization of many control parameters using
numerical algorithms based on Poyntriagin's Maximum principle (see
e.g.~\cite{JCP92p364,PRA61n012101,JCP118p8191,PRA63n032308, JCP120p5509,
JMR172p296}).  Optimal control may be the only viable option for
implementing quantum gates for systems with highly complex Hamiltonians
including off-resonant excitation and multi-body fixed coupling terms
~\cite{JMO56p831}, but numerical optimization can be time-consuming and
the resulting optimal control fields can be quite complicated and not
necessarily easy to implement.  By contrast, geometric control, vaguely
inspired by nuclear magnetic resonance (NMR)~\cite{CMR9p211}, requires
only sequences of simple pulses to implement arbitrary single and multi
qubit gates.  Although compared optimally designed pulses the results
may be suboptimal, this approach remains popular especially in an
experimental setting, due to its conceptual and experimental simplicity.
However, there are limits to the applicability of standard techniques
such as the Euler and Cartan decomposition, for instance, when we cannot
implement local rotations about orthogonal axes, a situation that
arises in various settings, from global electron-spin
achitectures~\cite{PRB72n045350} to charge-based semi-conductor quantum
dot systems~\cite{PRL95n090502}.

Geometric control generally relies heavily on Lie group decompositions
such as the standard Euler decomposition~\cite{Euler1862} of rotations
in $\RR^3$, which provides an explicit formula for decomposing any
rotation in $\RR^3$ into a sequence of (at most) three rotations about
two fixed, orthogonal axes, $\uvec{g}$ and $\uvec{h}$.  Due to the
equivalence of $\SU(2)$ and $\SO(3)$ ($\SO(3)\simeq\SU(2)/ \{-1,1\}$),
it also provides an explicit scheme to decompose any special unitary
operator in $\SU(2)$ into elementary complex rotations, combined with
the generalized Cartan decomposition for multi-qubit gates, it provides
a basis for generating arbitrary multi-qubit gates.  The main drawback
of the standard Euler angle decomposition is that requires orthogonal
rotation axes, or respectively, Hamiltonians, while the Hamiltonians
that are experimentally easily accessible are often at best
approximately orthogonal, subject to certain simplifications such as
negligible drift, rotating wave approximation, etc.  Applying the
standard Euler angle decomposition when the available basic Hamiltonians
are not orthogonal reduces the fidelity of most local gates, and hence
virtually all multi-qubit quantum gates, regardless of the quality of
the entangling gates, decoherence or other sources of noise that may
reduce the fidelities of quantum operations.  This is a significant
problem for applications such as quantum computation, where extremely
high accuracy of the elementary gates is a prerequisite for scalability.

One way to improve the accuracy of elementary gates is using composite
pulse sequences~\cite{JMR33p473,ProgNMR18p61,NJP2p6_1} to compensate for
certain systematic errors such as rotation axis alignment and rotation
angle errors.  Such approaches have proved to be extremely valuable in
ensemble-based quantum computing schemes such as liquid-state NMR, where
multiple qubits are encoded into different nuclear spins of a larger
molecule, and the system consists of an ensemble of a large number of
identical molecules in solution.  Due to magnetic field gradients,
diffusion processes and inter-molecular interactions, the actual fields
experienced by the individual molecules are subject to fluctuations,
resulting in rotation angle errors, and to a lesser extent, rotation
axis errors.  Composite pulses reduce these errors by replacing simple
unitary operations (rotations) with sequences of rotations designed to
``cancel'' certain errors.  However, this systematic error cancellation
comes at the expense of increased overhead in the number of elementary
operations, and hence time required to implement a single quantum gate,
especially if the systematic errors are so large as to require the use
of concatenated composite pulses~\cite{PhD_matt}.  This can exacerbate
other problems such as decoherence.  Composite pulses can be applied to
implement gates that are robust with respect to model uncertainty in
non-ensemble-based systems.  However, unlike in ensemble-based schemes,
where the systematic errors are a direct consequence of the fact that
different molecules in the ensemble experience different forces,
systematic errors due to model uncertainty in non-ensemble systems can
be minimized by experimental system identification~\cite{PRA69n050306,
PRA71n062312,PRA73n062333,PRA80n022333}, and this has been shown to be
advantageous in that it reduces to level of concatenation required for
composite pulse sequences~\cite{PhD_matt}.  In this paper we show that
if the actual Hamiltonians are known to sufficient accuracy then we can
significantly improve gate fidelities with minimal overhead simply by
optimizing the Euler angles in the decomposition, potentially completely
eliminating the need for expensive composite pulse sequences.

\section{Quantum Gate Engineering using Lie Group Decompositions}
\label{sec:gate-engineering}

Quantum computing generally relies on decomposing multi-qubit gates into
products of elementary single and two-qubit gates, which can be applied
simultaneously or sequentially to produce a desired unitary evolution.
Following the idea of using realistic physical Hamiltonians to generate
quantum gates efficiently~\cite{PRA71n052317,QCC4p93,AIPC963p748}, one
approach is to decompose a desired unitary operation $U$ into elementary
unitary operations that can be easily generated by natural Hamiltonian
flows of the system.  For instance, given a system with a natural Ising
coupling $\sz^{(1)}\sz^{(2)}$ between adjacent qubits, and the ability
to generate arbitrary local unitary operations, any two-qubit gate can
be factorized into a product of local operations, and the natural flows
$\op{Z}(t)=\exp(-i t\sz^{(1)}\sz^{(2)})$ using the Cartan decomposition
~\cite{PRL93n020502}
\begin{equation}
\label{eq:cartan}
  \op{U} = \op{U}_1 \,
           [\op{K}_x^\dag \, \op{Z}(\alpha_1) \op{K}_x]\,
           [\op{K}_y^\dag\, \op{Z}(\alpha_2) \, \op{K}_y]\,
           \op{Z}(\alpha_3) \, \op{U}_2,
\end{equation}
where $Z(\alpha)$ corresponds to free evolution of the system under the
Ising-coupling Hamiltonian for the time $t=\alpha$, and $\op{U}_1$,
$\op{U}_2$, $\op{K}_x$ and $\op{K}_y$ are simultaneous local operations
on both qubits.  $\op{U}_1$ and $\op{U}_2$ depend on the particular gate
to be implemented, while $\op{K}_x=\op{U}_x^{(1)}(\pi)\otimes
\op{U}_x^{(2)}(\pi)$, where
$\op{U}_x^{(k)}(\alpha)=\exp(-i\frac{\alpha}{2}\sx^{(k)})$, $k=1,2$, and
similarly for $\op{K}_y$.  The Cartan decomposition can be generalized
to interactions involving more than two qubits~\cite{PRA63n032308}, and
an explicit algorithm to calculate the generalized Cartan decomposition
was presented in~\cite{JMP46p001}.  Similar decompositions also exist
for other natural non-local Hamiltonians but we still require very
accurate single qubit gates.  In principle such gates are easy to
implement.  Any $W \in \SU(2)$ can be written as
\begin{equation}
\label{eq:W} W(\alpha,\beta,\gamma) = \begin{pmatrix}
    \cos(\alpha) \, {\rm e}^{ i \beta} &
    \sin(\alpha) \, {\rm e}^{ i \gamma} \\
   -\sin(\alpha) \, {\rm e}^{-i \gamma} &
    \cos(\alpha) \, {\rm e}^{-i \beta}
\end{pmatrix}
\end{equation}
where $0\leq \alpha \leq \frac{\pi}{2}$, $0\leq\beta,\gamma<2\pi$.  We
also have $W=\exp(-i\tilde{H})$ with $-i\tilde{H}\in\su(2)$, i.e.,
\begin{equation}
 \label{eq:H}
  \tilde{H} = \tilde{H}(\vec{d}) = d_x \sx + d_y \sy + d_z \sz
\end{equation}
with the usual Pauli matrices
\begin{equation}
 \sx = \begin{pmatrix} 0 & 1 \\ 1 & 0 \end{pmatrix}, \quad
 \sy = \begin{pmatrix} 0 & -i \\ i & 0 \end{pmatrix}, \quad
 \sz = \begin{pmatrix} 1 & 0 \\ 0 & -1 \end{pmatrix}.
\end{equation}
Let $\vec{d}=(d_x,d_y,d_z)$ with $\Omega=\norm{\vec{d}}$ and
$\vec{n}=\Omega^{-1}\vec{d}$.  As $H^2=\Omega^2 \Id$, where $\Id$ is the
identity matrix, we have
\begin{align*}
  e^{-i t H}
  &= \exp[-i \Omega t (n_x\sx+n_y\sy+n_z\sz)] \\
  &= \cos(\Omega t) \Id -i \sin(\Omega t) (n_x \sx + n_y \sy + n_z \sz)\\
  &= \begin{pmatrix}
      \cos(\Omega t)-i n_z \sin(\Omega t) & -(n_y+i n_x)\sin(\Omega t) \\
      (n_y-i n_x) \sin(\Omega t) & \cos(\Omega t)+i n_z \sin(\Omega t)
     \end{pmatrix}
\end{align*}
and comparison of the last equation with Eq.~(\ref{eq:W}) shows that
$W=\exp(-i T H)$ if we choose $\vec{n}$, $\Omega$ and $T$ such that
\begin{subequations}
\begin{align}
 \Omega T &= \arccos(\cos\alpha\cos\beta) \\
  \vec{n} &= -S^{-1} (\sin\alpha\sin\gamma,\sin\alpha\cos\gamma,\cos\alpha\sin\beta)
\end{align}
\end{subequations}
with $S=\sin(\Omega T)$.  Thus, if we have full control over the single
qubit Hamiltonians then we can implement any single qubit gate in a
single step, and if there are no constraints on the magnitude $\Omega$
of the Hamiltonian then the gate operation time $T$ can be made
arbitrarily small.

Unfortunately, for most physical systems we cannot implement arbitrary
Hamiltonians even locally.  For example, the single qubit Hamiltonians
for many potential qubit systems from ions to quantum dots are of the
form $H=\frac{1}{2}(d_z\sz+d_x\sx)$ or $H=\frac{1}{2}(d_x\sx+d_y\sz)$,
restricting us to rotations about axes in the $xz$ or $xy$ planes,
respectively.  If we have sufficient control over both $d_x$ and $d_z$
such as to be able to perform rotations about two orthogonal axes in the
plane, then we can still implement arbitrary single qubit gates using
the standard Euler decomposition, e.g.,
\begin{equation}
  W(\alpha,\beta,\gamma)
 = U_z\left(\beta+\gamma-\frac{\pi}{2}\right)
   U_x(\alpha) U_z\left(\beta-\gamma+\frac{\pi}{2}\right)
\end{equation}
where $U_x(\alpha)=\exp(i\frac{\alpha}{2}\sx)$ and
$U_z(\alpha)=\exp(i\frac{\alpha}{2}\sz)$ are elementary rotations about
the $x$ and $z$ axis respectively.

However, in practice there are often more constraints, limiting us to
varying one or both parameters within a certain range.  For example, for
certain solid-state architectures such as charge-based semi-conductor
quantum dot systems~\cite{PRL95n090502}, it is difficult or impossible
to dynamically control the tunnel coupling $d$ in the model Hamiltonian
$H=\Delta\omega\sz+d\sx$.  Thus $d\in [d_{\min},d_{\max}]$ and if
$d_{\min}>0$ then we cannot implement rotations about the $z$ axis, no
matter how much control we have over the energy level splitting
$\Delta\omega$, and practical constraints often make it impossible to
find operating parameters $(\Delta\omega_1,d_1)$ and
$(\Delta\omega_2,d_2)$ such that the corresponding Hamiltonians are
exactly orthogonal.  The same problem arises for other architectures
where the amount of control is limited, such as global electron spin
systems where many electron spins in quantum dots are simultaneously
controlled by a fixed global field, and we can only control the detuning
$\Delta\omega$ of individual spins from the global field via local
voltage gates~\cite{PRB72n045350}. In other cases $\Delta\omega$ may be
fixed while we have limited control over the coupling strength $d_x$ or
Rabi-frequency.

In these examples (and other similar systems) we have a fixed drift
Hamiltonian and constraints on a controllable parameter.  Without loss
of generality, let us consider $H(\kappa)=\frac{d}{2}(\sx+\kappa\sz)$
with $d>0$ fixed and $\kappa\in [0,\kappa_{\max}]$.  $\Tr(\sx\sz)=0$ and
$\Tr(\sx^2)=\Tr(\sz^2)=\Tr(\Id)=2$ shows that the Hilbert-Schmidt inner
product $\ip{H(\kappa_1)}{H(\kappa_2)}=\Tr[H(\kappa_1)^\dag H(\kappa_2)]$ 
satisfies
\begin{equation}
  \ip{H(\kappa_1)}{H(\kappa_2)}
  \le \ip{H(0)}{H(\kappa_{\max})}  = \frac{d^2}{2},
\end{equation}
and $\norm{H(\kappa)}=\sqrt{\ip{H(\kappa)}{H(\kappa)}}=d
\sqrt{(1+\kappa^2)/2}$.  Thus, provided $d\neq 0$, the angle
$\zeta$ between the Hamiltonians $H(0)$ and $H(\kappa_{\max})$ is
determined by
\begin{equation}
 \cos\zeta = \frac{\ip{H(0)}{H_{\kappa_{\max}}}}{
              \norm{H(0)}\cdot \norm{H_{\kappa_{\max}}}}
            = \frac{1}{\sqrt{1+\kappa_{\max}^2}}.
\end{equation}
Thus, $\zeta\to\frac{\pi}{2}$ only for $\kappa_{\max}\to\infty$.  For
any finite value of $\kappa_{\max}$ the maximum angle between the
accessible rotation axes will be less than $\frac{\pi}{2}$.  If we use
the standard Euler decomposition of a single qubit gate
\begin{equation}
  U = U_x(\alpha) U_z(\beta) U_x(\gamma),
\end{equation}
assuming $U_x(\alpha)=\exp(-i\frac{\alpha}{2}X)$ and
$U_z(\beta)=\exp(-i\frac{\beta}{2}Z)$, but the actual ``$z$''-rotation
is a rotation about $\tilde{Z}=X\cos\zeta +Z\sin\zeta$ with $\zeta=(1\pm
\eps)\frac{\pi}{2}$ then the gate actually implemented is
\begin{equation}
   \tilde{U} = U_x(\alpha) U_z^{\eps}(\beta) U_x(\gamma)
\end{equation}
with $U_z^\eps(\beta)=\exp(-i\frac{\beta}{2}\tilde{Z})$.  If there are
no other errors the gate fidelity will be
\begin{equation}
\begin{split}
  \F(\beta,\eps)
  &= \frac{1}{2}|\Tr(U^\dag\tilde{U})|
   = \frac{1}{2}|\Tr[U_z(\beta)^\dag U_z^\eps(\beta)]|\\
  &= \cos^2(\beta/2)[1-\cos(\eps\pi/2)] +|\cos(\eps\pi/2)|
\end{split}
\end{equation}
and the gate error $\E(\beta,\eps)=1-\F(\beta,\eps)$.  Thus the maximum
single qubit gate error is $1-|\cos(\eps\frac{\pi}{2})|=\E(\pm\pi,\eps)$
and, noting $\ave{\cos^2(x)}=\frac{1}{2}$, the average single qubit gate
error is $\E_{\rm avg}(\eps)=\frac{1}{2}[1-\cos(\eps\frac{\pi}{2})]$,
and for the maximum single qubit error to be below $10^{-4}$, the
rotation angle error must be less than $\eps=0.9$\% or equivalently
\begin{equation}
 \cos\left(\frac{\pi}{2}\eps\right) = \sin\zeta =
 \frac{\kappa_{\max}}{\sqrt{1+\kappa_{\max}^2}} \ge 0.9999
\end{equation}
or $\kappa_{\max}\ge 70.7054$.  Hence, to keep the maximum gate error
for a single qubit gate below the error threshold of $10^{-4}$, for
example, we would have to be able to make the energy splitting
$\Delta\omega$ (the controllable parameter) at least $71$ greater than
the fixed coupling $d$, even if there were no other sources of error. If
a CNOT-gate is implemented using the Cartan decomposition
(\ref{eq:cartan}) with $(\alpha_1,\alpha_2,\alpha_3)=(\pi/4,\pi/4,0)$
and $U_1=U_1^{(1)}\otimes U_1^{(2)}$, $U_2=U_2^{(1)}\otimes U_2^{(2)}$,
$K_x=K_x^{(1)}\otimes K_x^{(2)}$, $K_y=K_y^{(1)}\otimes K_y^{(2)}$,
where
\begin{subequations}
\label{eq:UCart}
\begin{align}
 K_x^{(1)}= K_x^{(2)}&= U_x(\pi) \\
 K_y^{(1)}= K_y^{(2)}&= U_x(\pi) U_z(\pi) \\
 U_1^{(1)}=U_2^{(1)} &= U_z(1.75\pi) \\
 U_1^{(2)}           &= U_x(0.5\pi) U_z(1.5\pi) U_x(1.5\pi) \\
 U_2^{(2)}           &= U_z(1.5\pi) U_x(0.5\pi)
\end{align}
\end{subequations}
then assuming that our $z$-rotations $U_z^\eps(\beta)$ are really
rotations about the tilted axis
$\tilde{Z}=X\sin(\frac{\pi}{2}\eps)+Z\cos(\frac{\pi}{2}\eps)$, shows
that the fidelity of the CNOT gate will be $<0.9999$ unless the rotation
axis angle error is less than about $0.6$\%, or $\kappa_{\max}\ge 100$,
even if the entanglement-generating Ising-coupling terms are perfect and
there are no other sources of error such as decoherence.  In practice,
other sources of error would mean that the error resulting from the
rotation axis misalignment would have to be much smaller, and thus
$\kappa_{\max}$ much bigger, for the total errors to remain below the
error threshold.  Also note that for $\kappa_{\max}=1$ the rotation axis
angle error is $50$\%, and the maximum single qubit gate error is
$1-\cos(\pi/4)$, almost $30$\%, and the error for a CNOT gate
implemented using the Cartan decomposition above with unoptimized single
qubit gates jumps to over $50$\%, assuming no errors in the Ising terms.

\section{Optimized Euler Decomposition}
\label{sec:opt_decomp}

The previous section shows that accurate single qubit gates are crucial,
and even small deviations of the experimentally accessible single qubit
Hamiltonians from orthogonality are problematic, not to mention
situations where the experimentally accessible Hamiltonians are far from
orthogonal.  It is also known that \emph{any} local unitary operation,
in principle, can be generated exactly by performing a sequence of
complex rotations about any two (fixed) Hamiltonians $H_1$ and $H_2$
that generate $\su(2)$, i.e., satisfy $[H_1,H_2]\neq 0$.  Various Lie
group decompositions have been considered for the related problem of
implementing local qubit operations exactly in the presence of various
types of fixed drift terms \cite{PRA62n053409}.  Ideally, however, we
would like a simple explicit algorithm to calculate an optimal sequence
of rotations given a fixed set of Hamiltonians (rotation axes) and an
arbitrary local gate.

In the following we consider general decompositions of $\SO(3)$ instead
of $\SU(2)$ using the equivalence between $\SU(2)$ and $\SO(3)$ (modulo
$\pm 1$).  The advantage of considering $\SO(3)$ is that it is easier to
visualize rotations in $\RR^3$ than complex rotations in $\SU(2)$.  As a
brief reminder we recall that any quantum state of a two-level system
can be represented by a density operator
\begin{equation}
\label{eq:rho}
  \rho = \rho(r,\theta,\phi) =\frac{1}{2}
 \begin{pmatrix}
    1+r^2\cos\theta & r^2e^{-i\phi}\sin\theta\\
    r^2e^{i\phi}\sin\theta & 1-r^2\cos\theta
 \end{pmatrix}
\end{equation}
with $0\le\theta\le\pi$, $0\le\phi\le 2\pi$ and $0\le r\le1$, and we can
define a unique mapping between density operators $\rho$ of constant purity
$\Tr[\rho^2]=\frac{1}{2}(1+r^2)$ and points on a sphere of radius $r$ in
$\RR^3$ by
\begin{equation}
\label{eq:s}
    \rho(r,\theta,\phi) \mapsto
    \vec{s}(r,\theta,\phi)= r
    \begin{pmatrix}
    \sin\theta\cos\phi \\
    \sin\theta\sin\phi \\
    \cos\theta
    \end{pmatrix}.
\end{equation}
The evolution of $\rho$ under a constant Hamiltonian $H$ then
corresponds to a rotation of the Bloch vector $\vec{s}$ about the unit
axis $\uvec{n}=\Omega^{-1}\vec{d}$ with the constant angular velocity
$\Omega=\norm{\vec{d}}$, as defined in Sec.~\ref{sec:gate-engineering}.
Given two Hamiltonians $H_1$ and $H_2$ we calculate the corresponding
normalized Bloch vectors $\uvec{h}$ and $\uvec{g}$ and note that the
angle between the rotation axes is given by
\begin{equation}
  \zeta=\arccos(\uvec{h}\cdot\uvec{g}),
\end{equation}
and $\zeta\neq0$ if and only if $[H_1,H_2] \neq 0$.

Any target operator $W(\alpha,\beta,\gamma)\in\SU(2)$ is equivalent
(modulo $\pm 1$) to a rotation $R(a,b,c)\in\SO(3)$ acting on the Bloch
vector $\vec{s}$ with $a=\frac{1}{2}\alpha$,
$b=\frac{1}{2}(\beta+\gamma)$ and $c=\frac{1}{2}(\beta-\gamma)$, and
we have explicitly
\begin{equation}
\label{eq:Rabc} R(a,b,c)=\begin{pmatrix}
    -b_1c_1+a_2b_2c_2 & b_2c_1+a_2b_1c_2 & -a_1c_2 \\
    -b_1c_2-a_2b_2c_1 & b_2c_2-a_2b_1c_1 &  a_1c_1 \\
     a_1b_2           & a_1b_1           & a_2
\end{pmatrix},
\end{equation}
where $a_1=\sin(a)$, $a_2=\cos(a)$, and similarly for $b$ and $c$.

\begin{algorithm}[h,t]
\caption{Calculate Generalized Euler Angles $\vec{\eps}$ for
 decomposition of arbitrary $R\in\SO(3)$.}
\label{algo2}
\begin{minipage}{\columnwidth}
\flushleft\parindent 0pt \textbf{Input:}
  $R\in\SO(3)$, unit vectors $\uvec{h},\uvec{g}\in \RR^3$,
  $\uvec{h}\neq\pm\uvec{g}$. \\
\textbf{Output:}  Euler angles $\vec{\eps}=(\eps_0,\ldots,\eps_{2p+2})$\\
 $\zeta =\arccos(\uvec{h}\cdot\uvec{g})$ \\
 $(\theta_{af},\phi_{af})=$
       \mbox{\textbf{Polar}}$(R\uvec{h},\uvec{h},\uvec{g})$\\
 $p=\ceil{\theta_{af}/2\zeta}$ \\
 \textbf{if} $p>0$ \\
 \QUAD $p=p-1$\\
 \textbf{end} \\
 $\theta = \theta_{af}-2\zeta p$ \\
 $\eps_1 = -\arccos[\cot\zeta \tan(\theta/2)]$ \\
 $\eps_2 = -\arccos[(-\cos^2\zeta+\cos\theta)/\sin^2\zeta]$\\
 $\eps_3 = \phi_{af}$ \\
 $T_p     = (\Rh(\pi)\Rg(\pi))^p$\\
 $S_1    = \Rg(-\eps_1)\Rh(-\eps_2) T_p \Rh(-\eps_3)$\\
 $(\theta_{b1},\phi_{b1})$
   =\mbox{\textbf{Polar}}$(S_1 R R_{\uvec{y}}(\pi/2)\uvec{h},\uvec{h},\uvec{g})$\\
 $\eps_{0}=\phi_{b1}$  \\
 \textbf{if} $p=3$ \\
  \QUAD $\eps = [\eps_{0},\eps_1,\eps_2+\eps_3]$\\
 \textbf{else} \\
  \QUAD $\eps = [\eps_{0},\eps_1,\eps_2+\pi,
         \underbrace{\pi,\ldots,\pi}_{2p-1},\eps_3]$\\
 \textbf{end} \\
 $\eps = \eps+2\pi\floor{(2\pi-\eps)/(2\pi)}$\\
\end{minipage}
\end{algorithm}
\begin{algorithm}[t]
\caption{\textbf{Polar}: Compute polar coordinates of $\uvec{a}$ with respect to
orthogonal frame induced by $\uvec{h}$ and $\uvec{g}$}
\label{algo0}
\begin{algorithmic}
\REQUIRE unit vectors $\uvec{a},\uvec{h},\uvec{g}\in \RR^3$
\ENSURE polar coordinates $(\theta,\phi)$ of $\uvec{a}$ \STATE
$\uvec{z} = \uvec{h}$ \STATE $\vec{y}  = \uvec{h}\times\uvec{g}$
\STATE $\uvec{y} = \vec{y}/\norm{\vec{y}}$ \STATE $\uvec{x} =
\uvec{y} \times\uvec{h}$ \STATE $\theta   =
\arccos(\uvec{a}\cdot\uvec{z})$ \STATE $\phi     =
\arctan(\uvec{a}\cdot\uvec{y},\uvec{a}\cdot\uvec{x})$
\end{algorithmic}
\end{algorithm}

\noindent{\bf Proposition.}  \emph{Any rotation $R\in\SO(3)$ about an
arbitrary axis in $\RR^3$ can be decomposed in a series of rotations
about two (non-identical) fixed rotation axes $\uvec{h}$ and $\uvec{g}$
in $\RR^3$ as follows:
\begin{equation}
\label{eq:euler3}
 R = \left\{ \begin{array}{ll}
 \Rh(\eps_3+\eps_2) \Rg(\eps_1) \Rh(\eps_0), & p=0\\
 \Rh(\eps_3) X^{p-1} \Rg(\pi) \Rh(\tilde{\eps}_2) \Rg(\eps_1) \Rh(\eps_0), &p>0
\end{array}\right. ,
\end{equation}
where the parameters $p$, $\eps_{0}$, $\eps_{1}$, $\eps_{2}$ and
$\eps_{3}$ are given explicitly in Algorithm~\ref{algo2}, and
$X=\Rg(\pi)\Rh(\pi)$, $\tilde{\eps}_2=\eps_2+\pi$.}

We would like emphasize here that the important part from an application
point of view is not the existence of a general decomposition of the
form~(\ref{eq:euler3}), which was predicted by \cite{72Lowenthal} and
shown in \cite{AUT40p1997}, but the simple Algorithm~\ref{algo2} to
compute the Euler angles $\mathbb{\eps}= (\eps_0,\ldots,\eps_{2p+2})$ in
the decomposition based on analytical formulas.  In most cases the Euler
angles in decompositions about non-orthogonal rotation axes can only be
determined numerically using optimization techniques~\cite{qph09071887},
but in this case we are in the fortunate position that we can derive
relatively simple analytic formulas for all Euler angles.

Before we discuss the derivation of this result and the algorithm, we
should briefly justify the use of the expression ``optimized'' Euler
angles in the title.  The factorization~(\ref{eq:euler3}) shows that in
general $2(p-1)+5=2p+3$ rotations are necessary, and for $\theta_f
\le\pi$ the maximum number of steps is
$2p+3\le\ceil{\frac{\pi}{\zeta}}+1$, which is equal to the order of
generation of $\SO(3)$, which is $k=\ceil{\frac{\pi}{\zeta}}+1$
according to Lowenthal's criterion~\cite{72Lowenthal}.  This means that
the decomposition is optimal in the sense that we achieve unit fidelity,
and that we cannot generate arbitrary rotations using rotations about
the fixed axes $\uvec{h}$ and $\uvec{g}$ in fewer steps in general~%
\footnote{The decomposition of individual gates is not always unique.
In some (rare) special cases the algorithm may produce a decomposition
that can be simplified further.  If we choose the orthogonal axes
$\uvec{h}=\uvec{z}=(0,0,1)^T$ and $\uvec{g}=\uvec{x}=(1,0,0)^T$, for
example, then Algorithm~\ref{algo2} returns
$\vec{\eps}=(\frac{3}{2}\pi,\pi,\frac{3}{2}\pi)$ for the swap gate
$\sx$, whose $\SO(3)$ representation is $R_X = \diag(1,-1,-1)$, which
corresponds to the decomposition $\Rh(\frac{3}{2}\pi)\Rg(\pi)
\Rh(\frac{3}{2}\pi)$ and can be simplified to $\Rg(\pi)$ as
$\Rz(a)\Rx(\pi)\Rz(a)=\Rx(\pi)$ for any $a\in\RR$.}.  
Optimality in terms of the number of rotation steps is often related to
time-optimality as more rotation steps generally will take longer to
complete, although time-optimality is not guaranteed.  For instance, if
rotations about one axis can be implemented much faster than rotations
about the other then a sequence that requires more steps but fewer slow
rotations by larger angles may be faster to implement.  Also, it should
be noted that while the decomposition~(\ref{eq:euler3}) generally
provides the best way to implement a quantum gate in $\SU(2)$ exactly
using a minimal number of rotations about two fixed axes, we may be able
to implement a particular gate substantially faster if we can
dynamically vary the rotation axes continuously.  For example, consider
a system with Hamiltonian $H=\frac{d}{2}(\sx+\kappa\sz)$.  If we can
temporally vary $\kappa(t)$ to take any value in the range
$[0,\kappa_{\max}]$ rather than two fixed values, e.g., $0$ and
$\kappa_{\max}$, then we may be able to implement a particular gate
faster by numerically optimizing $\kappa(t)$ instead of using the
optimized Euler angle decomposition.  The attractivity of the
generalized Euler decomposition lies in its simplicity.  It is a simple
``bang-bang'' control scheme that can be used to achieve unit fidelity
when we have limited control and cannot (or do not wish to) implement
complicated temporal control field profiles $\kappa(t)$.

The derivation of the algorithm is based on steering of a state
represented by a vector $\vec{s}_0\in \RR^3$ of length $r$ to another
state $\vec{s}_f\in\RR^3$ the same distance from the origin using only
rotations about the two fixed rotation axes given by the unit vectors
$\uvec{h}$ and $\uvec{g}$.  An algorithm and detailed explanation how to
move from one point on the sphere to another though a sequence of
rotations about two fixed axes $\uvec{h}$ and $\uvec{g}$ is presented in
Appendix~\ref{app:A}.  However, this algorithm on its own is not
sufficient for gate engineering as a single point and its image on the
unit sphere in $\RR^3$ are not sufficient to uniquely determine a
rotation in $R\in\SO(3)$.  Rather, we need at least two (non-antipodal)
points and their images to fix $R$.  This may seem surprising as the
parametrization~(\ref{eq:W}) for a unitary operator $W\in\SU(2)$ shows
that the image $W\ket{\psi}$ of a single Hilbert space vector
$\ket{\psi}$ is sufficient to fix all three parameters.  The
corresponding real rotation (\ref{eq:Rabc}), however, cannot be fully
determined by the image of a single Bloch vector.  Rather the mapping of
a single point in $\RR^3$ defines a one-parameter family of elements of
$\SO(3)$, and a second (non-antipodal) point and its image are required
to fix all three parameters $a,b,c$ of $R$.  The reason for this
apparent discrepancy is that when we transform from complex Hilbert
space vectors $\ket{\psi}$ to density matrices $\rho=
\ket{\psi}\bra{\psi}$ the information about the global phase of the
initial and final states, which helps fix $W$ uniquely, is lost.

Any pair of non-antipodal initial points $(\uvec{a}_0,\uvec{b}_0)$ and
their images $(\uvec{a}_f,\uvec{b}_f)$, are sufficient, but it is
convenient to choose the initial points to be the orthogonal set of 
initial states
\begin{subequations}
\begin{align}
    \uvec{a}_{0} &= \uvec{h} = (0,0), \\
    \uvec{b}_{0} &= \textstyle
                    R_{\uvec{y}}\left(\frac{\pi}{2}\right) \uvec{a}_0
                  = \left(\frac{\pi}{2},0\right),
\end{align}
\end{subequations}
where the pairs $(\theta,\phi)$ are the polar coordinates 
\begin{subequations}
\begin{align}
 \label{eq:thetarelspha}
 \theta &= \theta(\uvec{a}) = \arccos(\uvec{a}\cdot\uvec{z}) \\
 \label{eq:phirelspha}
  \phi &= \phi(\uvec{a}) = \arctan(\uvec{a}\cdot\uvec{y},\uvec{a}\cdot\uvec{x})
\end{align}
\end{subequations}
of the unit vectors $\vec{a}_0$ and $\vec{b}_0$ with respect to the
rectangular coordinate system $(\uvec{x},\uvec{y},\uvec{z})$ defined by
\begin{equation}
  \label{eq:coord}
  \uvec{z} = \uvec{h}, \quad
  \uvec{y} = \frac{\uvec{h}\times\uvec{g}}{\norm{\uvec{h}\times\uvec{g}}}, \quad
  \uvec{x} = \uvec{y}\times\uvec{h}.
\end{equation}
and the four-quadrand arctangent is
\begin{equation}
  \arctan(y,x) =
 \left\{ \begin{array}{ll}
  \arctan|y/x|      & x\ge 0,\, y\ge 0 \\
  \pi-\arctan|y/x|  & x<0,   \, y\ge 0 \\
  \pi+\arctan|y/x|  & x<0,   \, y<0 \\
  2\pi-\arctan|y/x| & x\ge 0,\, y<0.
  \end{array} \right.
\end{equation}

Then we compute the polar coordinates of the corresponding images under
the target rotation $R$
\begin{subequations}
\begin{align}
    \uvec{a}_{f}= R \uvec{a}_{0}=(\theta_{af},\phi_{af}), \\
    \uvec{b}_{f}= R \uvec{b}_{0}=(\theta_{bf},\phi_{bf}).
\end{align}
\end{subequations}
and use the state transfer algorithm (Algorithm~\ref{algo1}) to
calculate $S_1$, the series of rotations that steer $\uvec{a}_f$ to
$\uvec{a}_{0}$,
\begin{equation*}
  S_1 = \Rg(-\epsilon_1) \Rh(-\epsilon_2) T^{p} \Rh(-\phi_{af})
\end{equation*}
where $T=\Rh(\pi)\Rg(\pi)$ and $p$ is the largest integer strictly less
than $\theta_{af}/2\zeta$, i.e., $p=\ceil{\theta_{af}/2\zeta}-1$. The
generalized Euler angles $\eps_1$ and $\eps_2$ can be obtained by
inserting $\theta_f=0$ and $\theta_0=\theta_{af}-2p\zeta\neq 0$ into
subroutine PR2 (see Appendix~\ref{app:B})
\begin{subequations}
\label{eq:eps12}
\begin{align}
  \epsilon_2 &= -\arccos \left[
                \cot{\zeta} \tan\left( \frac{\theta_{af}-2p\zeta}{2} \right) \right] \\
  \epsilon_1 &= -\arccos \left[
                \frac{-\cos^2{\zeta}+\cos(\theta_{af}-2p\zeta)}{\sin^2{\zeta}}
                \right]
\end{align}
\end{subequations}
	
Since $S_{1}\uvec{b}_f=(\frac{\pi}{2},\phi_{b1})$ and
$\uvec{b}_0$ are unit vectors with the same $\theta$-angle
($\theta=\frac{\pi}{2}$), the same series of rotations
preserves the distance between the points, hence
\begin{align*}
  \uvec{b}_0 = \Rh(-\phi_{b1}) S_{1} \uvec{b}_{f}.
\end{align*}
The $\Rh$ rotations leave $\uvec{a}_{0}$ unchanged and we have
\begin{align*}
    \uvec{a}_0 = \Rh(-\phi_{b1}) S_{1} \uvec{a}_{f}.
\end{align*}
Thus $R=S_1^{-1}\Rh(\phi_{b1})$ and we have the decomposition
\begin{equation}
\label{eq:euler3a}
 R = \Rh(\eps_3) X^{p} \Rh(\eps_2) \Rg(\eps_1) \Rh(\eps_{0})
\end{equation}
where $X=T^{-1}=\Rg(\pi)\Rh(\pi)$ and $\eps_0=\phi_{b1}$,
$\eps_3=\phi_{af}$ and $\eps_1$, $\eps_2$ as in~(\ref{eq:eps12}).
If $p>1$ we can combine the two subsequent $\uvec{h}$ rotations,
while for $p=0$, $T^p=\Id$, thus the optimal decomposition is
\begin{equation*}
 R = \left\{ \begin{array}{ll}
 \Rh(\eps_3+\eps_2) \Rg(\eps_1) \Rh(\eps_0), & p=0\\
 \Rh(\eps_3) X^{p-1} \Rg(\pi) \Rh(\tilde{\eps}_2) \Rg(\eps_1) \Rh(\eps_0), &p>0
\end{array}\right.
\end{equation*}
with $\tilde{\eps}_2=\eps_2+\pi$, which completes the proof.


%

\section{Applications}

To apply the results in the previous section to implement a
quantum gate
\begin{equation}
  U = \exp[i \Phi (n_x \sx + n_y \sy + n_z\sz)]
\end{equation}
given the Hamiltonians
\begin{equation}
  H_1 = \frac{d}{2}\sx, \quad
  H_2 = \frac{d}{2}(\sx + \kappa \sz),
\end{equation}
we identify the normalized Hamiltonians $\tilde{H}_{1}=\sx$ and
$\tilde{H}_2=(\sx+\kappa\sz)/\sqrt{1+\kappa^2}$ with the unit
vectors $\uvec{h}=(1,0,0)^T$ and
$\uvec{g}=(1,0,\kappa)^T/\sqrt{1+\kappa^2}$, respectively, and use
Algorithm~\ref{algo2} to decompose the corresponding
$\SO(3)$-representation of the target operator $U$
\begin{equation}
  A = \exp[\Phi (n_x R_x + n_y R_y + n_z R_z)]
\end{equation}
where $R_x$, $R_y$ and $R_z$ are the rotation generators
\begin{equation*}
 R_x = \begin{pmatrix}
         0 & 0 & 0 \\
         0 & 0 & 2 \\
         0 &-2 & 0
       \end{pmatrix},
 R_y = \begin{pmatrix}
         0 & 0 & -2 \\
         0 & 0 & 0  \\
         2 & 0 & 0
       \end{pmatrix},
 R_z = \begin{pmatrix}
         0 & 2 & 0 \\
        -2 & 0 & 0 \\
         0 & 0 & 0
       \end{pmatrix}.
\end{equation*}

Table~\ref{table1} shows the optimized Euler angle decomposition
results for the gates $S=\exp\left(i\frac{\pi}{4}\sz\right)$,
$T=\exp\left(i\frac{\pi}{8}\sz\right)$,
\begin{equation} \textstyle
  U_{\rm Had} = \exp\left(i\frac{\pi}{2\sqrt{2}}(\sx+\sz)\right)
\end{equation}
as well as for the single qubit gates~(\ref{eq:UCart}) required to
implement a CNOT gate via the Cartan
decomposition~(\ref{eq:cartan}). Gates that require only $\sx$
rotations have been omitted as they are trivial to implement with
the given Hamiltonians.  Using the optimized Euler angles rounded
to four significant digits, the gate errors for all single qubit
gates in the table, as well as the CNOT gate, are below $3\times
10^{-9}$ for values of $\kappa$ shown, while the gate errors using
the standard Euler angles increase to almost $30$\% for $K_y$ and
$\kappa=1$.  The error for the resulting CNOT gate increases from
$\approx 10^{-4}$ for $\kappa=100$ to over $51$\% for $\kappa=1$.
Also note that the penalty for non-orthogonal Hamiltonians in
terms of the number of rotation steps required is actually rather
small unless $\kappa$ is very small.  Indeed for $\kappa\ge 1$ all
of the elementary gates in the table can be implemented in at most
four steps, and for $\kappa>1$, this is indeed the maximum number
of steps required for any single qubit gate.  To see this recall
that Lowenthal's criterion guarantees that the maximum number of
steps in the decomposition of any single qubit gates is
\begin{equation}
 K= \left\lceil \frac{\pi}{\zeta} \right\rceil +1 =
   \left\lceil\frac{\pi}{\arccos[(1+\kappa^2)^{-1/2}]}\right\rceil +1,
\end{equation}
which yields $K=3$ for $\kappa=\infty$ and $K=4$ for
$1<\kappa<\infty$, $K=5$ for $\kappa=1$, $K=6$ for
$\sqrt{[\cos(\pi/5)]^{-2}-1}<\kappa<1$, and so forth.

\begin{table}
\begin{tabular}{|l||l||c||c|c|c|c|}
\hline
  & $\kappa$ & $\E_0$ (\%)
             & $\eps (\tilde{H}_1)$
             & $\eps (\tilde{H}_2)$
             & $\eps (\tilde{H}_1)$
             & $\eps (\tilde{H}_2)$ \\\hline
$T$ & $\infty$ &  0   &       0 &  1.7500 &  0     &\\
 & 100      &  0.0007 &  0.0013 &  1.7500 &  0.0013&\\
 &  50      &  0.0029 &  0.0026 &  1.7499 &  0.0026&\\
 &  10      &  0.0727 &  0.0132 &  1.7487 &  0.0132&\\
 &   5      &  0.2844 &  0.0264 &  1.7448 &  0.0264&\\
 &   1      &  4.2893 &  0.1359 &  1.6359 &  0.1359&\\
\hline
$S$ & $\infty$ &  0      &  0      &  1.5000 &  0     &\\
 & 100      &  0.0025 &  0.0032 &  1.5000 &  0.0032&\\
 &  50      &  0.0100 &  0.0064 &  1.4999 &  0.0064&\\
 &  10      &  0.2481 &  0.0319 &  1.4968 &  0.0319&\\
 &   5      &  0.9710 &  0.0641 &  1.4873 &  0.0641&\\
 &   1      & 14.6446 &  0.5000 &  1      &  0.5000&\\
\hline 
$U_{\rm Had}$ & $\infty$ &  0      &  1.5000 &  1.5000 &  1.5000&\\
 & 100      &  0.0025 &  1.5032 &  1.5000 &  1.5032&\\
 &  50      &  0.0100 &  1.5064 &  1.4999 &  1.5064&\\
 &  10      &  0.2481 &  1.5319 &  1.4968 &  1.5319&\\
 &   5      &  0.9709 &  1.5641 &  1.4873 &  1.5641&\\
 &   1      & 14.6442 &  0      &  1      & 0      &\\
\hline 
$U_1^{(2)}$ &$\infty$ &  0      &  0.5000 &  1.5000 &  1.5000&\\
 & 100      &  0.0025 &  0.5032 &  1.5000 &  1.5032&\\
 &  50      &  0.0100 &  0.5064 &  1.4999 &  1.5064&\\
 &  10      &  0.2481 &  0.5319 &  1.4968 &  1.5319&\\
 &   5      &  0.9709 &  0.5641 &  1.4873 &  1.5641&\\
 &   1      & 14.6443 &  1      &  1      &  0     &\\
\hline 
$U_2^{(2)}$ &  $\infty$&  0      &  0      &  1.5000 &  0.5000&\\
 & 100      &  0.0025 &  0.0032 &  1.5000 &  0.5032&\\
 &  50      &  0.0100 &  0.0064 &  1.4999 &  0.5064&\\
 &  10      &  0.2481 &  0.0319 &  1.4968 &  0.5319&\\
 &   5      &  0.9709 &  0.0641 &  1.4873 &  0.5641&\\
 &   1      & 14.6445 &  0.5000 &  1      &  1     &\\
\hline 
$K_y^{(1)}$ & $\infty$ &  0      &  0      &  0      &  1      &  1\\
 & 100      &  0.0050 &  0.5000 &  1.9936 &  0.5000 &  1\\
 & 50       &  0.0200 &  0.5001 &  1.9873 &  0.5001 &  1\\
 & 10       &  0.4963 &  0.5032 &  1.9362 &  0.5032 &  1\\
 &  5       &  1.9419 &  0.5127 &  1.8718 &  0.5127 &  1\\
 &  1       & 29.2893 &  1      &  1      &  1      &  1\\
\hline
\end{tabular}
\caption{Optimized Euler angles (in units of $\pi$) for various
single qubit gates and different values of $\kappa$.
$\eps(\tilde{H}_1)$ indicates a rotation by $\eps$ about the
normalized axis $\tilde{H}_1$. $\E_0$ is the gate error that
results if the standard Euler angles for $\kappa=\infty$ are used.
The gate errors using the optimized Euler angles truncated to four
decimal digits are $<3\times 10^{-9}$ for all gates, and can be
made arbitrarily small by increasing the number of significant
digits of the Euler angles.}\label{table1}
\end{table}


\section{Conclusions}

The Euler decomposition of unitary operators in $\SU(2)$ is widely
used to implement single qubit gates by decomposing them into
products of rotations about two orthogonal axes determined by
fixed Hamiltonians. The approach can be problematic however as
experimentally accessible Hamiltonians in many cases may not be
orthogonal.  Depending on the situation, in some cases the
Hamiltonians can be made almost orthogonal, while in others the
constraints may be far more severe.  In either case, however, lack
of orthogonality of the underlying Hamiltonians leads to errors in
the gates implemented, and even small errors can propagate.  A
rotation axis angle error of even $1$\% results in single qubit
gate errors above the error threshold of $10^{-4}$ even if there
are no other sources of errors, and the single qubit errors
compound and lead to even larger errors for two-qubit gates.  Such
systematic errors can easily be corrected, however, by adapting
the Euler decomposition to the actual Hamiltonians available.

We have presented a explicit algorithm to calculate the optimized
Euler angles for any single qubit gate and two arbitrary fixed
Hamiltonians, and shown that we can substantially improve single
and two-qubit gate fidelities by using optimized rather than
standard Euler angles.  The idea is attractive because the
computational overhead to calculate the optimized Euler angles is
minimal and the implementation is no more demanding than standard
geometric control, i.e., no additional resources are required.
There is a small price to pay in terms of an increase in the
number of rotation steps required to implement a particular gate,
but unless the maximum angle between the experimentally accessible
Hamiltonians is very small, this increase is very slight, e.g.,
from at most three steps for orthogonal Hamiltonians to four for
Hamiltonians with angle $\zeta$ greater than $45^\circ$ and five
if $\zeta=45^\circ$. For a model Hamiltonian
$H(\kappa)=\frac{d}{2}(\sx+\kappa\sz)$ with a fixed coupling
parameter $d$, this condition is satisfied if the energy level
splitting can be made at least as large as the tunnelling energy
$d$, or $\kappa=1$, whereas the standard Euler decomposition would
require $\kappa\to\infty$, or energy level splittings that are
orders of magnitude greater than the tunnelling energy $d$ to
achieve near-orthogonal Hamiltonians.

Overall, the overhead in terms of complexity of the pulse
sequences is small compared to alternative ways to correct for
rotation axis errors, such as composite pulse sequences, and this
overhead seems acceptable, considering that relaxing the need to
be able to perform rotations about orthogonal axes may allow for
substantial simplifications of the underlying architectures.
Another source of overhead of the technique is the need for
initial characterization of the Hamiltonians.  It must also be
stressed that optimized Euler angles are designed to minimize
errors for a single system.  They cannot compensate for ensemble
errors, i.e., errors arising from the fact that individual systems
in a large ensemble may experience different fields and thus
different effective rotations.  However, the approach is an
effective way to improve gate fidelities for non-ensemble systems
with non-ideal Hamiltonians.

Further work is necessary to extend the results to
higher-dimensional systems.  Another issue is that different gates
require different amounts of time to implement.  This is not a
problem for a single system but would be for a large register if
one wants to implement gates on different qubits simultaneously.
Here we have only used two fixed Hamiltonians with a fixed angle
between them.  In many cases, however, we may be able to vary the
controllable parameter continuously up to some maximum value.  An
interesting question in this regard is whether we can exploit the
(limited) variation in the tilt angle to design simple geometric
controls that allow us to implement arbitrary gates in a fixed
amount of time.

\section{Acknowledgements}

SGS acknowledges support from EPSRC ARF Grant EP/D07195X/1, European
Union Knowledge Transfer Programme MTDK-CT-2004-509223, Hitachi and NSF
Grant PHY05-51164.  GC is supported by Grant VZ3407391 of the Ministry
of Education, Youth and Sports of the Czech Republic. CD acknowledges
support of the Program of the Cultural Exchanges-2008 between the Czech
and the Hellenic Republic.

\appendix

\section{State transfer algorithm}
\label{app:A}

\begin{figure}[t]
\scalebox{0.4}{\includegraphics{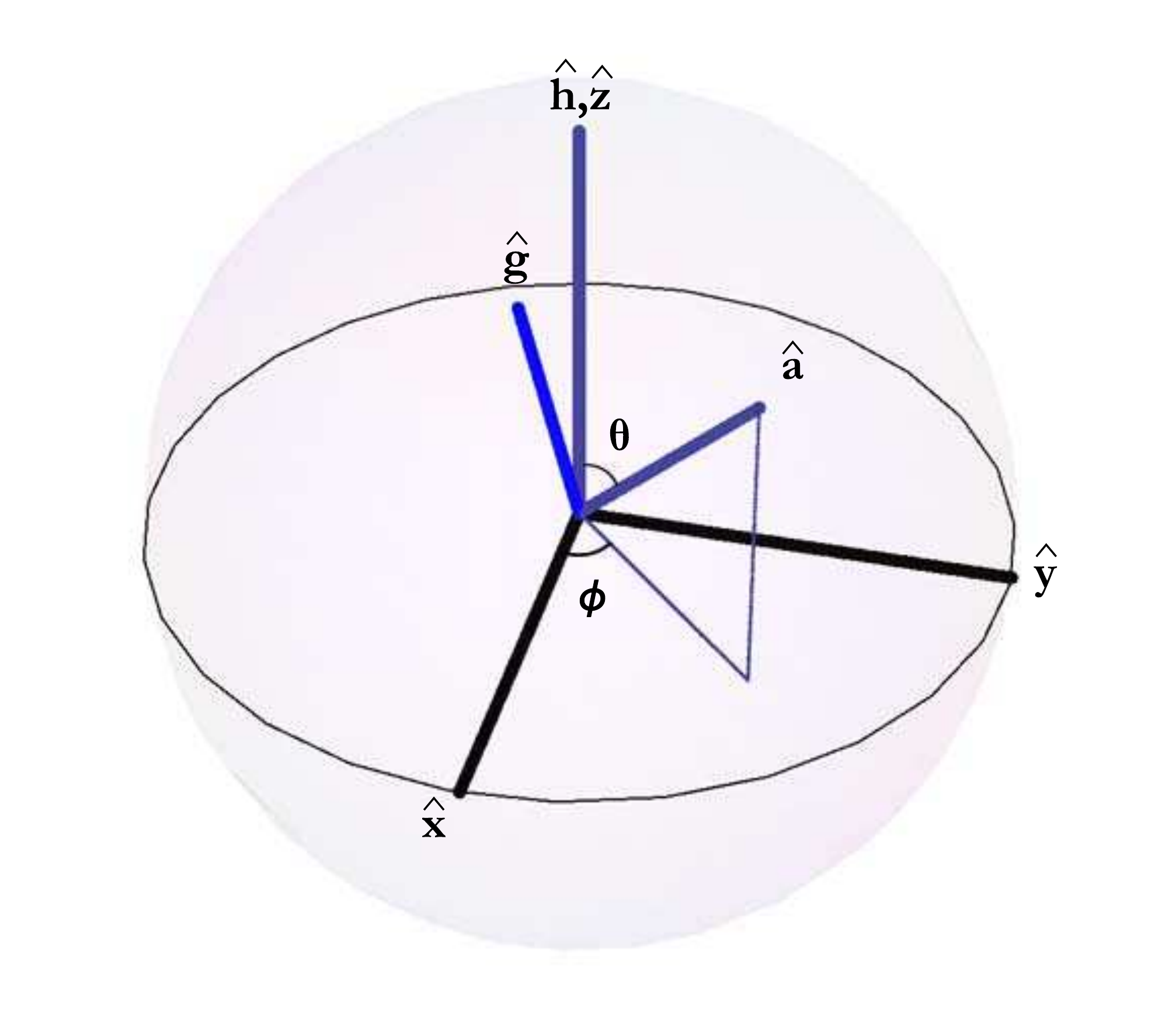}} \caption{Sphere with
arbitrary rotation axes, respective coordinate system and an
arbitrary vector $\uvec{a}$ (angles $\theta$ and $\phi$).}
\label{fig1}
\end{figure}

The objective of state transfer is to steer the system from a known
initial state $\vec{s}_0=(r,\theta_0,\phi_0)$ on a sphere of radius $r$
to a target state $\vec{s}_f=(r',\theta_f,\phi_f)$.  With unitary
control only states on the same sphere as the initial state are
accessible by performing a sequence of rotations about the axes
$\uvec{h}$ and $\uvec{g}$, respectively.  We shall assume $r=1$, noting
that the sequence of rotations that steers the normalized initial state
$\uvec{a}_0$ to the normalized target state $\uvec{a}_f$ steers
$\vec{s}_0$ to $\vec{s}_f$ if they lie on the same sphere of radius $r$,
and to a state $\vec{s}_f'$ that is as close to the target state as we
can get with unitary control if $r\neq r'$.  

A general strategy to get from $\vec{s}_0$ to $\vec{s}_1$ with a minimum
number of rotations about the axes $\uvec{h}=\uvec{z}$ and
$\uvec{g}=(\sin\zeta,0,\cos\zeta)^T$, following~\cite{AUT40p1997} and
earlier work~\cite{72Lowenthal,SCL2p341, RMJM21p879}, is to rotate the
initial state by a suitable angle about either axis to map it to a point
on the great circle in the $\uvec{x},\uvec{z}$ plane, followed by a
sequence of $\pi$-rotations, alternating about the $\uvec{h}$ and
$\uvec{g}$ axis, until the angle $\theta'$ of the current state differs
by less than $2\zeta$ from the $\theta_f$ of the target state, i.e., we
are within direct reach of the target state, followed by a final
rotation by a suitable angle about the same axis we started with.  By
Lowenthal's criterion any state can be reached from any other state in
at most $k+1$ steps, $k$ being the smallest integer
$\ge\frac{\pi}{\zeta}-1$~\cite{72Lowenthal}, and for
$\zeta=\frac{\pi}{2}$ at most two steps are required.

Based on this idea, we can derive an explicit algorithm for calculating
the generalized Euler angles of an optimal decomposition given the angle
$\aa$ between the rotation axes $\uvec{h}$ and $\uvec{g}$, and the
relative coordinates $(\theta_0,\phi_0)$ and $(\theta_f,\phi_f)$ of the
initial and final state.  Assume $\theta_0>\theta_f$ and
$\zeta<\pi/2$. If $\theta_0-\theta_f\leq 2\zeta$ then we can get from
$(\theta_0,0)$ to $(\theta_f,0)$ in two steps, either by rotating
$(\theta_0,0)$ around $\uvec{g}$ by an angle $\theta$, followed by a
rotation about $\uvec{h}$ by an angle $\phi$ (Subroutine PR1), or by a
rotation around $\uvec{h}$ by an angle $\phi$, followed by a rotation
around $\uvec{g}$ by an angle $\theta$ (Subroutine PR2).  If
$\theta_0-\theta_f> 2\zeta$ then we move from the initial point to a
point with $\theta_0'-\theta_f\leq 2\zeta$ via a sequence of
$\pi$-rotations around axes $\uvec{g}$ and $\uvec{h}$ as described
before.  If $\zeta\leq \pi/2$ but $\theta_0<\theta_f$ then we exchange
the initial and final points, apply the algorithm and finally reverse
the sequence of rotations.  If $\zeta>\pi/2$ we set
$(\theta_{1},\phi_{1})=(\pi-\theta_0,\phi_0)$,
$(\theta_{2},\phi_{2})=(\pi-\theta_f,\phi_f)$ and
$\tilde{\zeta}=\pi-\zeta$ and apply the algorithm.

The algorithm returns a list of pairs $(\eps,\uvec{r})$, where $\eps\in
[0,\pi]$ is a generalized Euler angle and $\uvec{r}=\uvec{h}$ or
$\uvec{r}=\uvec{g}$ indicates the rotation axis, which defines the
necessary sequence of the rotations.  E.g.  if $\theta_0>\theta_f$,
$\zeta<\frac{\pi}{2}$ and routine PR1 was used, then
\begin{equation}
   \vec{s}(\theta_f,\phi_f) =
   \Rh(\phi_f)\Rh(\phi)\Rg(\theta)T^{p}\Rh(-\phi_0) \vec{s}(\theta_0,\phi_0).
\end{equation}
where the angles $\phi$ and $\theta$ are given by
\begin{subequations}
\begin{align}
  \cos\phi&=\frac{\sin\theta_0-\cot{\zeta}(\cos\theta_f-\cos\theta_0)}{\sin\theta_f}\\
  \cos\theta&= \frac{-\cos\zeta\cos(\zeta-\theta_0)+\cos\theta_{f}}
                    {\sin\zeta\sin(\zeta-\theta_0)}.
\end{align}
\end{subequations}
Similarly, if PR2 was used
\begin{equation}
   \vec{s}(\theta_f,\phi_f) = \Rh(\phi_f)\Rg(\theta)\Rh(\phi)
   T^{p}\Rh(-\phi_0)\vec{s}(\theta_0,\phi_0)
\end{equation}
where the angles $\phi$ and $\theta$ are determined by
\begin{subequations}
\begin{align}
  \cos\phi  &=\frac{\sin\theta_f-\cot{\zeta}(\cos\theta_0-\cos\theta_f)}
                   {\sin\theta_0}\\
  \cos\theta&=\frac{-\cos{\zeta}\cos(\zeta-\theta_f)+\cos\theta_0}
                   {\sin{\zeta}\sin(\zeta-\theta_f)}.
\end{align}
\end{subequations}
The procedures PR1 and PR2 are described in Appendix~\ref{app:B},
and in both cases we have
\begin{equation}
    T^{p}=\underbrace{\Rh(\pi)\Rg(\pi)\ldots
                      \Rh(\pi)\Rg(\pi)}_{\mbox{2p rotations}}.
\end{equation}



\begin{algorithm}[t]
\caption{Calculate sequence of rotations about arbitrary, fixed
axes $\uvec{h}$ and $\uvec{g}$ required to move from one point on
the unit sphere to another.} \label{algo1}

\begin{minipage}{\columnwidth}
\flushleft\parindent 0pt \textbf{Input:} $(\theta_0,\phi_0)$,
$(\theta_f,\phi_f)$ -- polar coordinates of initial and final
point with respect to relative coordinate
system~(\ref{eq:coord}).\\
\textbf{Output:} List of pairs $(\eps,\uvec{u})$, $\uvec{u}\in
\{\uvec{h},\uvec{g}\}$ defining rotation steps necessary to get
from initial state to final state.\\
\textbf{if} $\theta_0=\theta_f$
       \textbf{return} $\{(\phi_f-\phi_0,\uvec{h})\}$\\
\textbf{else}\\
   \QUAD $p=\floor{(\theta_0-\theta_f)/2\zeta}$ \\
   \QUAD $\theta_0=\theta_0-2p\zeta$ \\
   \QUAD \textbf{if} $\theta_0=\theta_f$ \textbf{return}
       \[
    \{(\phi_f,\uvec{h}),
                 \underbrace{(\pi,\uvec{g}),(\pi,\uvec{h}),\ldots
                             (\pi,\uvec{g}),(\pi,\uvec{h})}_{\mbox{2$p$ pairs}},
                         (-\phi_0,\uvec{h}) \}
       \]
   \QUAD\textbf{else if}
         $\theta_0\le\zeta$ or ($2\zeta-\theta_0\ge 0$ and
            $2\zeta-\theta_0>\theta_f$) \\
   \QUAD\QUAD\textbf{return}
        \begin{align*}
            \{(\phi_f,\uvec{h}),\underbrace{(\pi,\uvec{g}),
             (\pi,\uvec{h}), \ldots,(\pi,\uvec{g}), (\pi,\uvec{h})}_{\mbox{2p pairs}},\\
            {\rm PR2}[\theta_0,\theta_f,\zeta],(-\phi_0,\uvec{h})\}
           \end{align*}
   \QUAD\QUAD\textbf{else} \textbf{return}
            \begin{align*}
             \{(\phi_f,\uvec{h}),\underbrace{(\pi,\uvec{g}),
                (\pi,\uvec{h}), \ldots,(\pi,\uvec{g}), (\pi,\uvec{h})}_{\mbox{2p pairs}},\\
             {\rm PR1}[\theta_0,\theta_f,\zeta],(-\phi_0,\uvec{h})\}
            \end{align*}
   \QUAD\QUAD\textbf{end} \\
   \QUAD\textbf{end} \\
\textbf{end}
\end{minipage}
\end{algorithm}

The minimal number of steps can be calculated explicitly.  If $\theta_0
=\theta_f$ then only a single rotation about $\uvec{h}$ is required,
otherwise the minimal number of steps to get from $(\theta_0,\phi_0)$ to
$(\theta_f,\phi_f)$ starting with a rotation about $\uvec{h}$ is
$N=2p+2+\Theta(q)$, where $\Theta(q)=1$ for $q>0$ and $0$ otherwise,
\begin{subequations}
\begin{align}
 p &= \mbox{\rm int}[(\theta_0 - \theta_f)/2 \zeta] \\
 q &= \cos\theta_f- \left[\tan\zeta(\sin\bar{\theta}-\sin\theta_f\cos\phi_f)
      +\cos\bar{\theta}\, \right]
\end{align}
\end{subequations}
and $\bar{\theta}=\theta_0-2p\zeta$.  Here, ${\rm int}(x)$ indicates the
integer part of $x$.

The minimal number of steps to get from $(\theta_0,\phi_0)$ to
$(\theta_f,\phi_f)$ starting with a rotation about $\uvec{g}$ is
$N'=2+\Theta(\theta'-\theta_f)\left[2p+1+\Theta(q)\right]$, where
\begin{subequations}
\begin{align}
 p &= \mbox{int}[(\theta'-\theta_f)/2 \zeta] \\
 q &= \cos\theta_f- \left[\tan\zeta(\sin\bar{\theta}-\sin\theta_f\cos \phi_f)
     +\cos\bar{\theta}\, \right]
\end{align}
\end{subequations}
with
$\theta'=\arccos(\cos\zeta\cos\theta_0+\sin\zeta\sin\theta_0\cos\phi_0)-\zeta$
and $\bar{\theta}=\theta'-2 p\zeta$.


\section{Subroutines PR1 and PR2}
\label{app:B}

Both procedures take only the $\theta$-angles of the initial and
final points, $\theta_0$ and $\theta_{f}$, respectively, and the
angle $\aa$ between the axes $\uvec{h}$ and $\uvec{v}$ as input,
assuming that the points have already been shifted to the
$\uvec{x}-\uvec{z}$-plane.

\subsection*{Subroutine PR1}

\begin{figure}
\scalebox{0.4}{\includegraphics{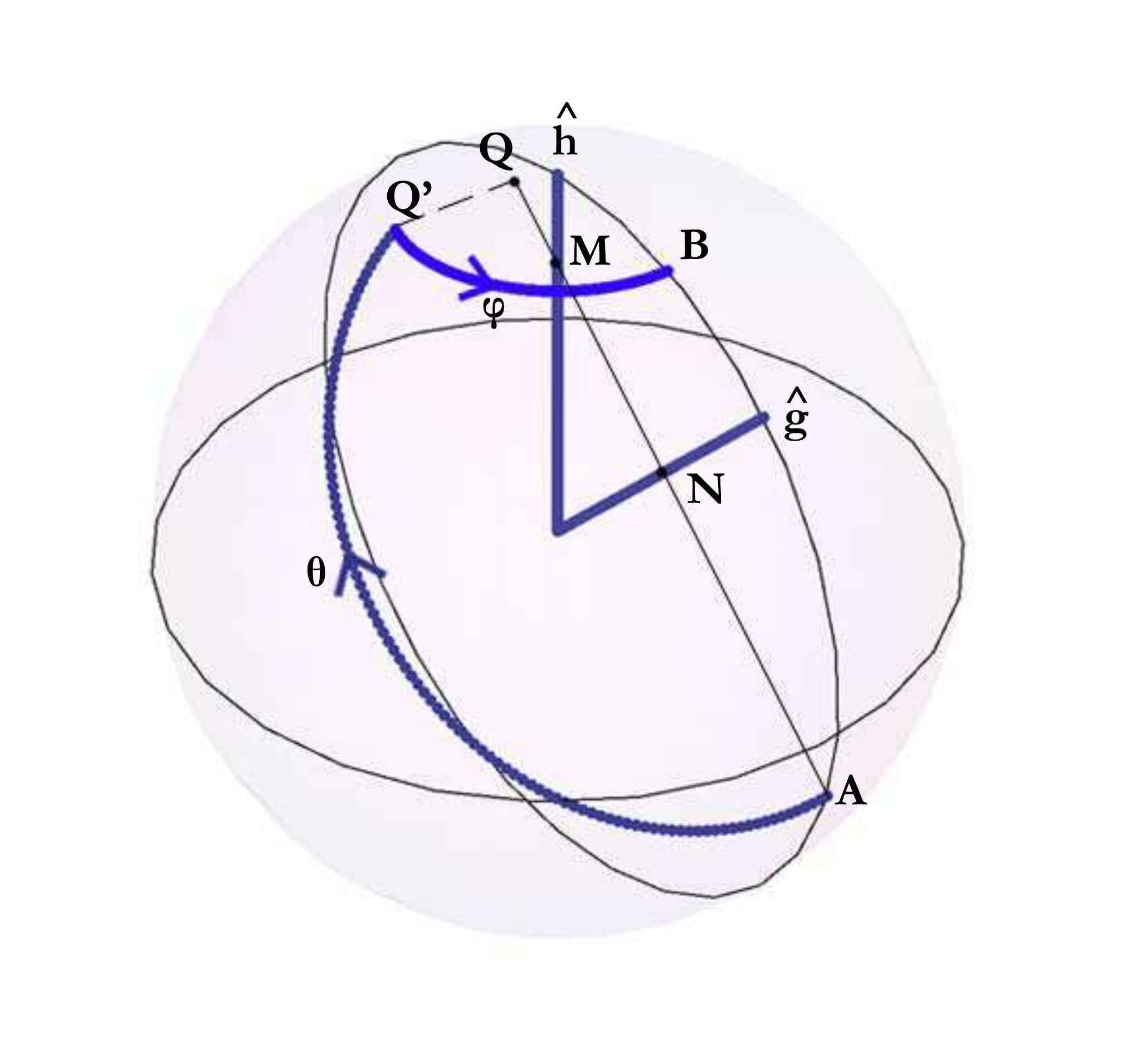}}
\scalebox{0.5}{\includegraphics{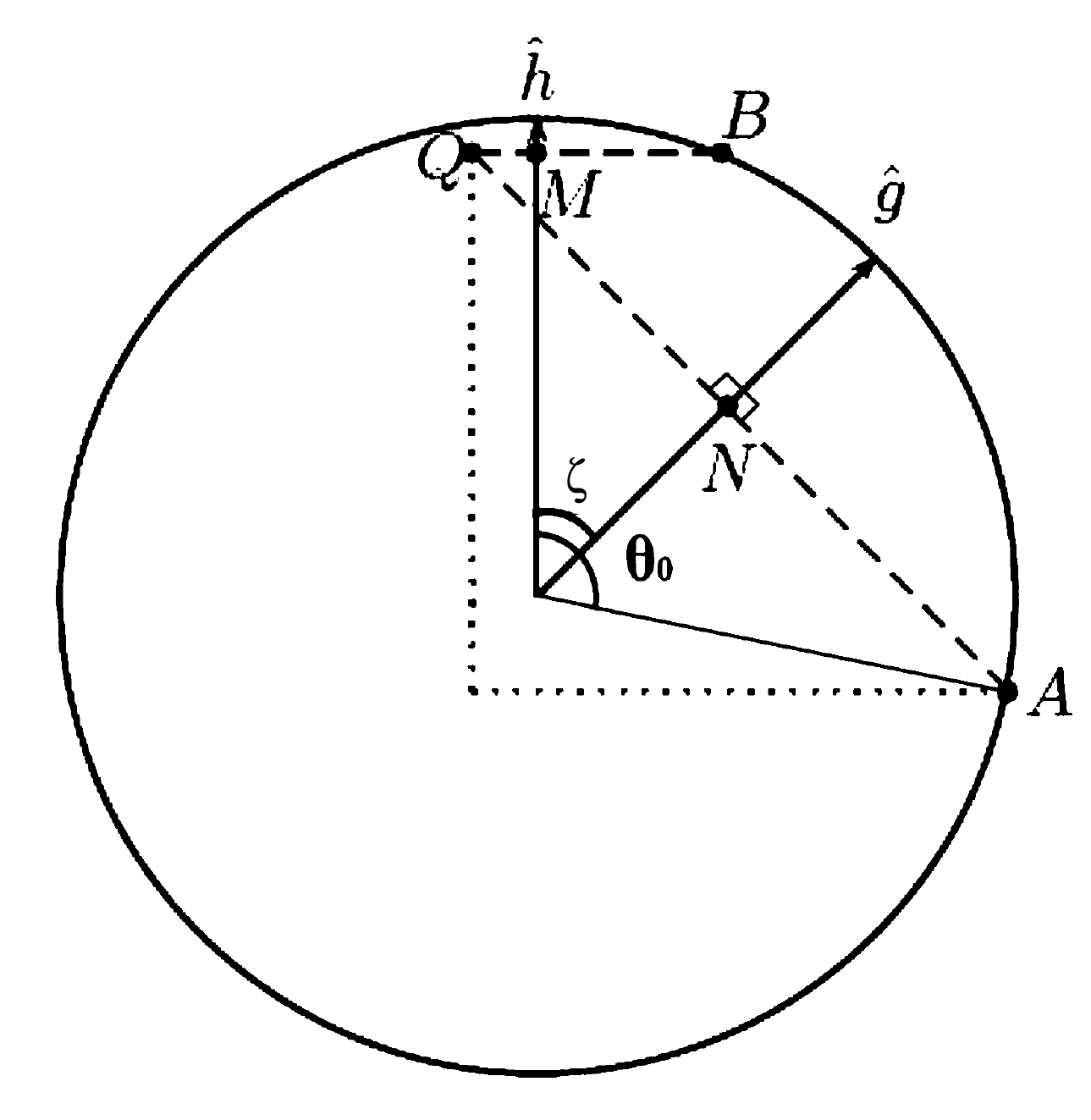}}
 \caption{Rotations on the sphere and projection to the $x-z$ plane for
 subroutine PR1}
 \label{fig:PR1}
\end{figure}

Fig.~\ref{fig:PR1} shows that $\norm{\vec{MB}}=\norm{\vec{MQ'}}$
and $\norm{\vec{NA}}=\norm{\vec{NQ'}}$ and
\begin{equation}
 \cos{\phi}  = \frac{|\vec{MB} \cdot \vec{MQ'}|}{\norm{\vec{MB}}^2}, \quad
 \cos{\theta}= \frac{|\vec{NA} \cdot \vec{NQ'}|}{\norm{\vec{NA}}^2}.
\end{equation}
Noting that $\vec{A}=(\sin\theta_0,0,\cos\theta_0)$ and
$\vec{Q}'=(q,p,\cos\theta_f)$ for suitable values of $p$ and $q$,
and taking $\vec{Q}=(q,0,\cos\theta_f)$ to be the projection of
$\vec{Q}'$ onto the $\uvec{x}\uvec{z}$ plane, shows
\begin{align}
   \cot{\zeta}
   = -\frac{\vec{QA}\cdot\uvec{x}}{\vec{QA}\cdot\uvec{z}}
   = \frac{\sin\theta_0-q}{\cos\theta_0-\cos\theta_f}
\end{align}
and thus $q=\sin\theta_0-\cot{\zeta}(\cos\theta_f-\cos\theta_0)$.
Noting further that $\vec{B}=(\sin\theta_f,0,\cos\theta_f)$ and
$\vec{M} =(0,0,\cos\theta_f)$, shows that $\vec{MQ'}=(q,p,0)$,
$\vec{MB}=(\sin\theta_f,0,0)$, and therefore
\begin{equation}
  \cos\phi = \frac{q \sin\theta_f}{\sin^2\theta_f}
           = \frac{\sin\theta_0-\cot{\zeta}(\cos\theta_f-\cos\theta_0)}{\sin\theta_f}.
\end{equation}

Furthermore, we have $\vec{N}=r(\sin\zeta,0,\cos\zeta)$ with
$r=\cos(\theta_0-\zeta)$ and $|\vec{NA}| = \sin(\theta_0-\zeta)$,
and
\begin{align*}
 \vec{NA} &=(\sin\theta_0-r\sin\zeta,0,\cos\theta_0-r\cos\zeta)\\
 \vec{NQ'}&=(q-r\sin\zeta,p,\cos\theta_0-r\cos\zeta)\\
 \vec{NA}\cdot\vec{NQ'} &=
 (\sin\theta_0-n_x)(q-n_x)+(\cos\theta_0-n_z)^2,
\end{align*}
which after some simplification gives
\begin{equation}
 \cos\theta = \frac{-\cos{\zeta}\cos{(\zeta-\theta_0)}+\cos{\theta_{f}}}
                   { \sin{\zeta}\sin{(\zeta-\theta_0)}}.
\end{equation}

\subsection*{Subroutine PR2}

\begin{figure}
\scalebox{0.4}{\includegraphics{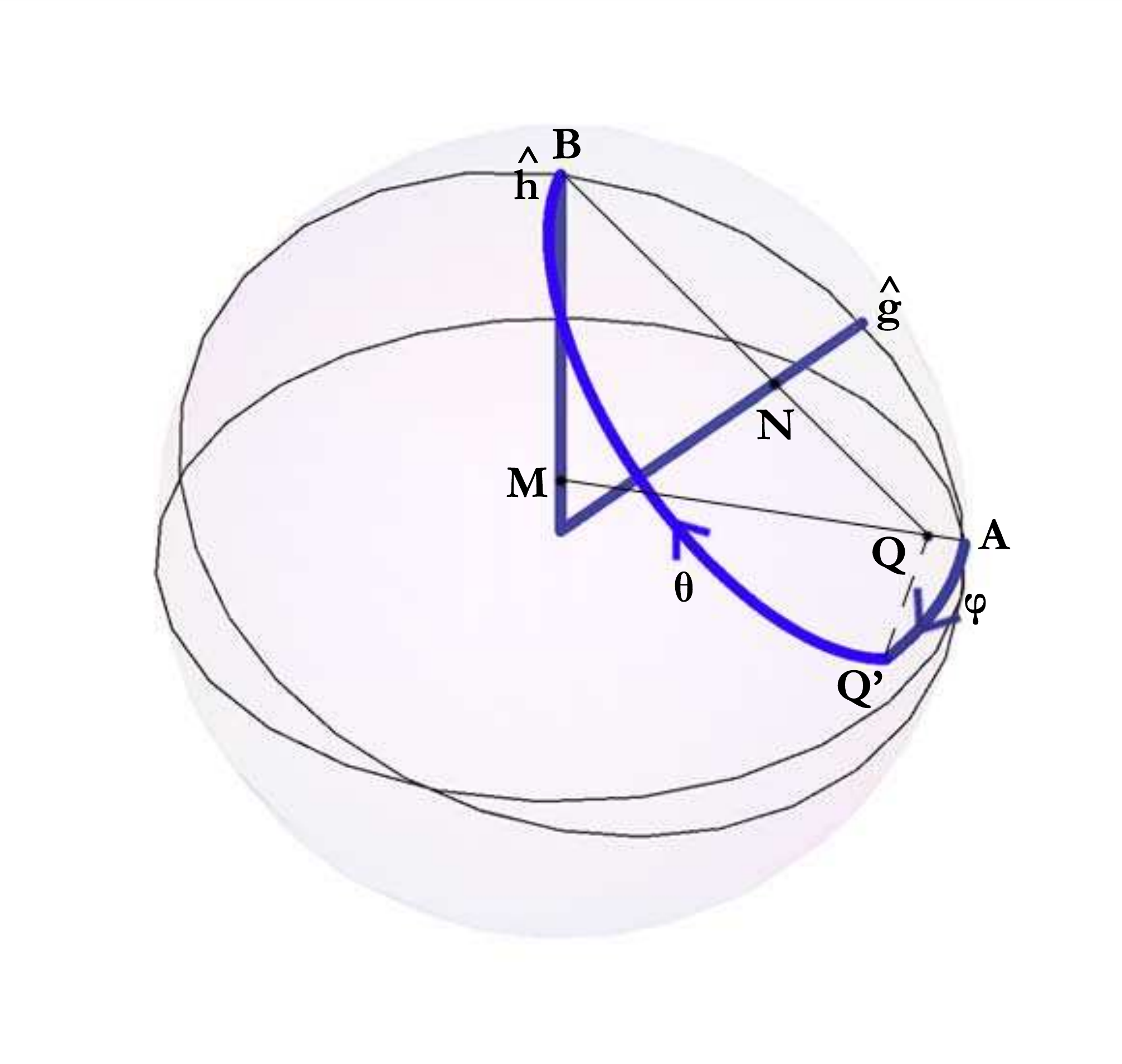}}
\scalebox{0.5}{\includegraphics{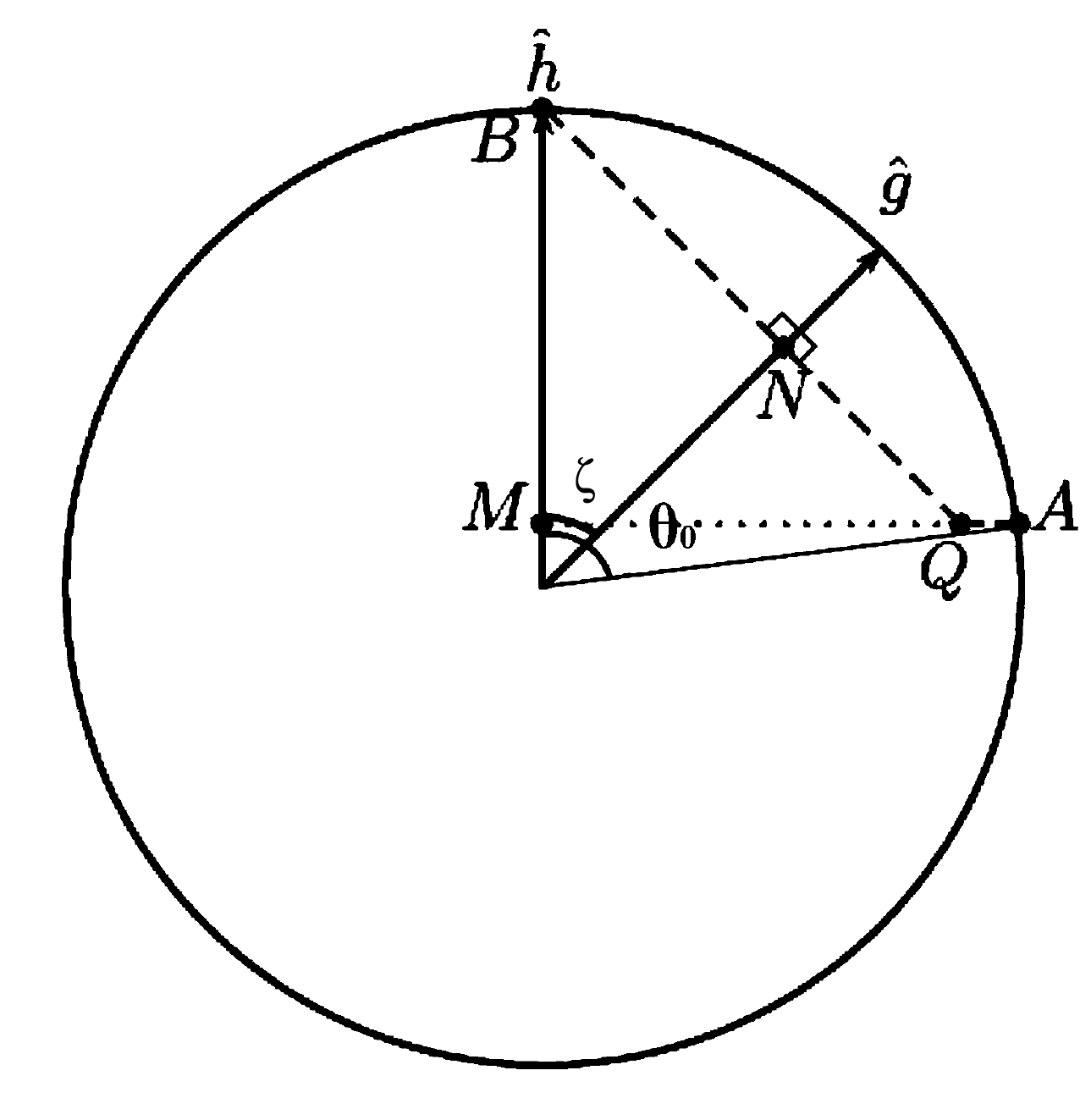}}
 \caption{Rotations on the sphere and projection to the $x-z$ plane for
 subroutine PR2}
 \label{fig:PR2}
\end{figure}

Fig.~\ref{fig:PR2} shows that $\norm{\vec{MA}}=\norm{\vec{MQ'}}$
and $\norm{\vec{NB}}=\norm{\vec{NQ'}}$ and
\begin{align}
    \cos \phi   = \frac{|\vec{MA}||\vec{MQ'}|}{|\vec{MA}|^2}, \quad
    \cos \theta = \frac{|\vec{NB}||\vec{NQ'}|}{|\vec{NB}|^2}.
\end{align}
Noting that $\vec{B}=(\sin\theta_f,0,\cos\theta_f)$ and
$\vec{Q}'=(q,p,\cos\theta_0)$ for suitable $p$ and $q$ as before,
shows that
\begin{align}
    \cot{\zeta}=
    -\frac{\vec{QB}\cdot\uvec{x}}{\vec{QB}\cdot\uvec{z}}=
    -\frac{\sin \theta_f-q}{\cos\theta_f-\cos\theta_0}
\end{align}
i.e., $q=\sin\theta_f-\cot{\zeta}(\cos\theta_0-\cos\theta_f)$.

Taking $\vec{Q}=(q,0,\cos\theta_0)$ to be the projection of
$\vec{Q}'$ onto the $\uvec{x}\uvec{z}$ plane, and noting that
$\vec{A} =(\sin\theta_0,0,\cos\theta_0)$ and
$\vec{M}=(0,0,\cos\theta_0)$ shows that $\vec{MQ'}=(q,p,0)$,
$\vec{MA}=(\sin\theta_f,0,0)$, and thus
\begin{equation}
  \cos\phi = \frac{q \sin\theta_0}{\sin^2\theta_0}
           = \frac{\sin\theta_f-\cot{\zeta}(\cos\theta_0-\cos\theta_f)}{\sin\theta_0}.
\end{equation}

Furthermore, we have $\vec{N}=r(\sin\zeta,0,\cos\zeta)$ with
$r=\cos(\theta_f-\zeta)$ and $|\vec{NB}| = \sin(\theta_f-\zeta)$,
and
\begin{align*}
 \vec{NB} &=(\sin\theta_f-r\sin\zeta,0,\cos\theta_f-r\cos\zeta)\\
 \vec{NQ'}&=(q-r\sin\zeta,p,\cos\theta_0-r\cos\zeta)\\
 \vec{NA}\cdot\vec{NQ'} &=
 (\sin\theta_0-n_x)(q-n_x)+(\cos\theta_0-n_z)^2,
\end{align*}
which after some simplification gives
\begin{align}
    \cos\theta
    &= -\cot{\zeta}\cot(\zeta-\theta_f)+ \cos\theta_0\csc\zeta\csc(\zeta-\theta_f) \nonumber\\
    &= \frac{-\cos{\zeta}\cos(\zeta-\theta_f)+\cos\theta_0}{\sin{\zeta}\sin(\zeta-\theta_f)}.
\end{align}



\bibliography{/home/sonia/archive/bibliography/References}
\bibliographystyle{prsty}

\end{document}